\documentclass[prl,twocolumn,twoside,superscriptaddress,floatfix]{revtex4}
\usepackage{amsmath,graphicx,multirow,hyperref}
\hypersetup{colorlinks,citecolor=black,filecolor=black,linkcolor=black,urlcolor=black}

\begin{document}

\title{Encouraging moderation:  Clues from a simple model of ideological conflict}

\author{Seth A. Marvel}
\email{smarvel@umich.edu}
\affiliation{Department of Mathematics, University of Michigan, Ann Arbor, Michigan 48109, USA}

\author{Hyunsuk Hong}
\affiliation{Department of Physics and Research Institute of Physics and Chemistry, Chonbuk National University, Jeonju 561-756, Korea}

\author{Anna Papush}
\affiliation{Department of Mathematics, Cornell University, Ithaca, New York 14853, USA}

\author{Steven H. Strogatz}
\affiliation{Department of Mathematics, Cornell University, Ithaca, New York 14853, USA} 

\date{\today}

\begin{abstract}
Some of the most pivotal moments in intellectual history occur when a new ideology sweeps through a society, supplanting an established system of beliefs in a rapid revolution of thought.  Yet in many cases the new ideology is as extreme as the old.  Why is it then that moderate positions so rarely prevail?  Here, in the context of a simple model of opinion spreading, we test seven plausible strategies for deradicalizing a society and find that only one of them significantly expands the moderate subpopulation without risking its extinction in the process.
\end{abstract}

\maketitle

The social history of ideas involves the frequent replay of a single story:  there is a widely accepted and deeply ingrained dogma in the community.  This dogma helps to justify the community's institutions and shape its common practices.  Then, in the midst of this stable milieu, a new doctrine emerges.  Backed by a small group of unwavering advocates, it challenges the status quo and steadily wins converts, eventually replacing the previous system to become the dominant ideology of the group.

In some cases, there is an enduring consensus that the new doctrine marks a tangible improvement on the old. This is the case for the American civil rights movement~\cite{morris84}, women's suffrage~\cite{flexner96,ramirez97}, and paradigm shifts in science~\cite{kuhn96,chen04,bettencourt06}.  However in many other situations, the newer doctrine is not clearly better. After some time as the dominant approach, it too is overtaken by a younger alternative, which in turn is itself replaced, and so on.  This second situation is often seen in rapidly spreading political campaigns~\cite{gerring01}, the booms and busts of credit lending and consumer confidence~\cite{dellariccia06,ludvigson04}, cultural fashions and short-lived reforms (e.g.,~Prohibition in the United States)~\cite{sproles81,bikhchandani92}, methodological or topical fads in academia, and various political revolutions~\cite{goldstone01}.

A natural question is this:  why do communities, and sometimes entire societies, get caught in these swings from one ideological extreme to the other when neither delivers a sustainable solution?  Why does not a majority of the population settle on an intermediate position that blends the best of the old and new?

\begin{figure}[t!]
\includegraphics[scale = 1]{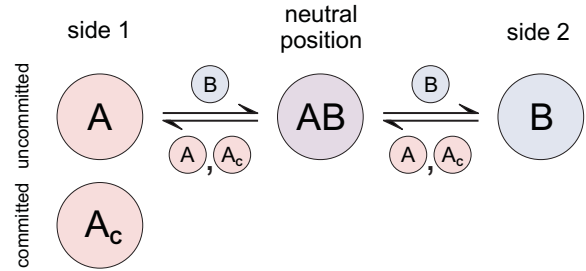}
\caption{Model structure (see text for definitions of $A$, $B$, $AB$, and $A_c$).  Labels on the arrows indicate the allowed affiliation(s) of a speaker that converts a listener from one subpopulation to another in the direction of the arrow. 
\label{ModelStructure}}
\end{figure}

There are several ways in which this question might be answered, but here we give one that is purely mathematical:  the environment of successive ideological revolutions is not conducive to moderate mindedness simply from a \textit{dynamical} perspective.  In particular, almost all of the intuitive ways of encouraging moderation either fail to expand the moderate subpopulation or make it vulnerable to collapse in the process of encouraging its growth.

In this Letter, we provide evidence for this claim by studying a minimal model of ideological revolution.  Critically, this model only addresses large-scale ideological conversions and does not treat the many other common processes found in real communities, such as apparent conversions within the old paradigm and situations where there is no conversion at all but rather a splitting of opinions, or fragmentation.  

The model (Fig.~\ref{ModelStructure}) starts from an assumption of a community consisting of four nonoverlapping subpopulations:  those that currently hold an extreme opinion $A$; those that currently hold the conceptually opposing opinion $B$ (in the simplest case, just the negation of $A$); those that currently hold neither $A$ nor $B$ (the moderates); and those that hold $A$ indefinitely and are immune to the influence of others (we call these committed believers or $A$ zealots).  We partially overload notation, using $A$, $B$, $AB$, and $A_c$, respectively, to denote both the individuals in these four subpopulations and the subpopulations themselves.  This model builds on earlier pioneering work in sociophysics~\cite{vazquez03, vazquez04, centola05, gekle05, delalama06, castello06, galam07, castellano09} and is directly inspired by (but different from) a model examined in a recent study of opinion dynamics~\cite{xie11,difference}.

\begin{table}[t!] 
\caption[Table 1]{Interactions that change the membership of subpopulations $A$, $B$, and $AB$ in the basic model.  The $A_c$ subpopulation is constant.} 
\label{table1}
\begin{tabular*}{\columnwidth}{@{\extracolsep{\fill}} ccc}
\hline
\hline
\multirow{2}{*}{Speaker} & Listener & Listener \\
                         & pre-interaction & post-interaction \\
\hline
\multirow{2}{*}{$A,A_c$} & $B$      & $AB$     \\ \cline{2-3}
                         & $AB$     & $A$      \\ \hline
\multirow{2}{*}{$B$}     & $A$      & $AB$     \\ \cline{2-3}
                         & $AB$     & $B$      \\ 
\hline
\hline
\end{tabular*}
\end{table}

The dynamics of the basic model are deterministic, continuous and mean-field, derived as the large-population limit of the following random process:  time is discrete, and at each time step we select two individuals uniformly at random and randomly choose one of them to be the \textit{speaker} and the other the \textit{listener}.  If the speaker is an $A$ or $B$ and the listener is a $B$ or $A$, respectively, then the listener is dissuaded from his or her extremist position and becomes an $AB$.  However, if the listener is an $AB$, then the listener becomes an $A$ if the speaker is an $A$ and a $B$ if the speaker is a $B$.  In all other cases, there is no change in the state of the speaker and listener (Table~\ref{table1}).  Note that in this highly simplified model, moderate speakers do not produce a change of mind in either their listeners or themselves; only extremists successfully rally others to their cause.

Let $n_A$, $n_B$, and $n_{AB}$ denote the expected fractions of the total population of $N$ individuals corresponding to the uncommitted $A$, $B$, and $AB$ subpopulations, respectively, and let $p$ denote the constant fraction of the population in the committed $A_c$ subpopulation. We will study how varying $p$, the proportion of zealots, affects the eventual state of the system.  Using this new notation, we can consider the expected change to the subpopulation fractions in the limits of a large population and a vanishing time step (which we take to grow like $N$ and shrink like $1/N$, respectively).  This reduces our discrete dynamics to the following rate equations:
	\begin{equation}
	\begin{split}
	\dot{n}_A &= (p+n_A) n_{AB} - n_A n_B, \\
	\dot{n}_B &= n_B n_{AB} - (p+n_A) n_B,
	\end{split}
	\label{eq1}
	\end{equation}
where $n_{AB} = 1 - p - n_A - n_B$ and the overdot denotes differentiation by time.  Since we present no formal evidence that the dynamics of (\ref{eq1}) do actually occur in practice, our work could alternatively be viewed as posing this model and its subsequent generalizations as interesting in their own right.

\begin{figure}[t]
\includegraphics[scale = 1]{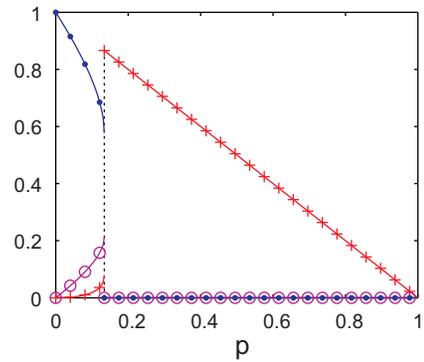}
\caption{The equilibrium values of $n_A$ (red plus signs), $n_B$ (blue dots), and $n_{AB}$ (magenta open circles) for the basic model as functions of $p$, assuming an initial population with $(n_A,n_B) = (0,1-p)$.  The vertical dashed line marks the critical value $p_c = 1-\sqrt{3}/2 \approx 0.134$.  At values of $p$ greater than $p_c$, $n_B$, and $n_{AB}$ are zero and $n_A = 1-p$.
\label{BasicModel}}
\end{figure}

Now suppose we run the system (\ref{eq1}) to equilibrium starting from a population initially composed of only $A_c$ and $B$ individuals. We will use this initial condition for all the systems considered in this Letter; the idea is that $A$ represents the new doctrine and $B$ the reigning view.  If we then track the final fractions of $n_A$, $n_B$, and $n_{AB}$ as functions of $p$, we find (as in Ref.~\cite{xie11}) that the equilibrium state changes dramatically as we increase $p$ through a critical value $p_c$ (Fig.~\ref{BasicModel}).  For $p < p_c$ the system remains similar to how it started---most of the individuals maintain $B$.  However as $p$ is increased through $p_c$, the system undergoes a discontinuous transition, and for $p > p_c$ the entire population quickly reaches a consensus on $A$.  A bifurcation analysis shows that $p_c = 1 - \sqrt{3}/2 \approx 0.134$~\cite{suppmat}.

\begin{figure}[b!]
\includegraphics[scale = 1]{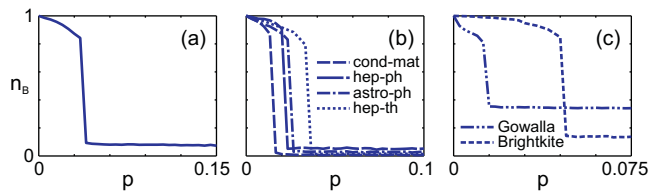}
\caption{The equilibrium fraction of $B$ believers remaining after the basic model is run on the giant connected component of (a) the U.S.~network of corporate board memberships in 1994~\cite{davis96}, (b) four coauthorship networks in the physics division of \href{arXiv.org}{arXiv.org}~\cite{leskovec07}, and (c) the friendship networks of the location-based social networking websites Gowalla and Brightkite~\cite{cho11}.  Note the abrupt transitions in $n_B$; compare with the corresponding curve in Fig.~\ref{BasicModel}.
\label{BasicModelOnRealNetworks}}
\end{figure}

To test the robustness of these mean-field predictions, we simulate the model on a diverse set of real social networks.  Figure~\ref{BasicModelOnRealNetworks} shows that in each case, the $n_B$ vs $p$ curves resemble the mean-field result depicted in Fig.~\ref{BasicModel}.  The primary differences are a lower $p_c$ value for the real networks and a small, stable fraction of peripherally located $B$ believers for $p > p_c$.

\begin{figure*}[t!]
\includegraphics[scale = 1]{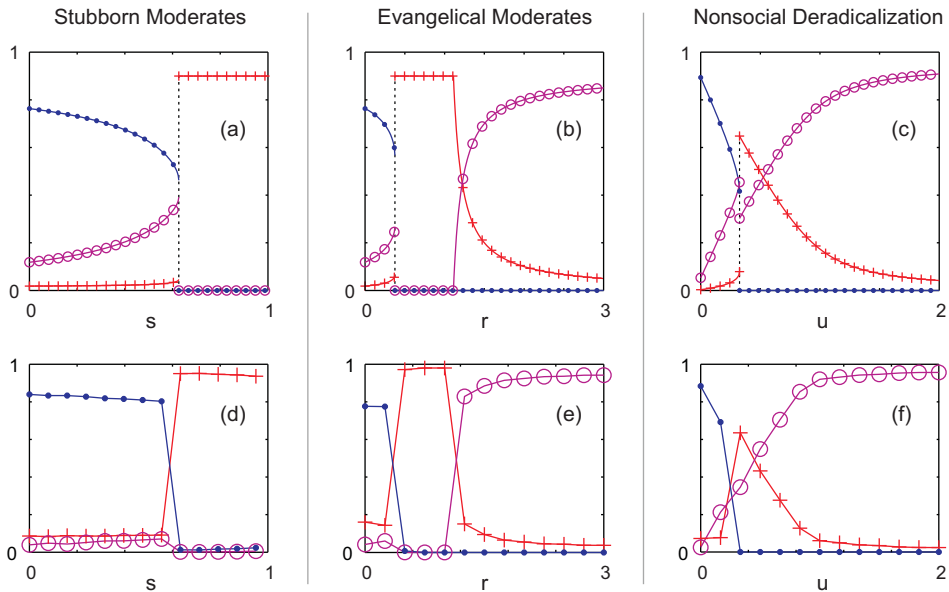}
\caption{(a)-(c):~Mean-field results obtained analytically for generalizations of the basic model with (a) stubborn moderates, (b) evangelical moderates, and (c) nonsocial deradicalization (see the text for details). The final equilibrium values of $n_A$ (red plus signs), $n_B$ (blue dots), and $n_{AB}$ (magenta open circles) for the initial condition $(n_A,n_B) = (0,1-p)$ are plotted as a function of the new parameter ($s$, $r$, or $u$) in the corresponding generalized model. Of the three strategies shown---and in fact for all seven considered in the Supplemental Material~\cite{suppmat}---only nonsocial deradicalization allows for the growth of the moderate fraction up to $1-p$ without risking its extinction.
(d)-(f):~Representative simulation results for the discrete-time versions of the models with (d) stubborn moderates, (e) evangelical moderates, and (f) nonsocial deradicalization when these models are run on the \href{arXiv.org}{arXiv} coauthorship network for high energy physics theory (hep-th)~\cite{leskovec07}.  The plots show the equilibrium fractions of $n_A$ (red plus signs), $n_B$ (blue dots), and $n_{AB}$ (magenta open circles) obtained.  Each simulation is started from the state in which a random but highly interconnected fraction $p$ of the population is committed to a belief in $A$ and the rest believe $B$.  The simulation is then run for $10^8$ time steps after which the values of $n_A$, $n_B$, and $n_{AB}$ are tabulated.  The constant fractions of zealots in the six panels of this figure are (a)~0.1, (b)~0.1, (c)~0.05, (d)~0.035, (e)~0.02, and (f)~0.02.
\label{GeneralizationsOnRealNetworks}}
\end{figure*}

With (\ref{eq1}) as our starting point, we now ask how we might alter the model to encourage moderation.  Specifically, we would like to (i)~increase the equilibrium size of the moderate subpopulation, and (ii)~decrease the chance that this equilibrium size could drop substantially if the parameter values (just $p$ for the basic model) were to vary a little.  In search of a strategy that does both, we explore seven different generalizations of the basic model.  Three generalizations are discussed here in the main text; the rest are treated in the Supplemental Material~\cite{suppmat}.  Figures~\ref{GeneralizationsOnRealNetworks}(a)-(c) summarize mean-field results for these three generalizations, and for comparison, the corresponding simulation results on a real social network are shown in the panels beneath them [Figs.~\ref{GeneralizationsOnRealNetworks}(d)-(f)].  Importantly, these do not constitute full empirical validations of the model and its generalizations (which would require dynamical data that are hard to obtain).  Rather, we include these simulations only as an indication of where the results of such tests might lie.  Furthermore, we only consider the equilibrium values reached from the pre-revolution initial condition $(n_A,n_B) = (0,1-p)$; as dynamical systems, the basic model and its generalizations are capable of a wider range of behaviors~\cite{suppmat}.

As a first attempt at achieving (i) and (ii) above, suppose we could somehow make the moderates less likely to convert to either of the two radical positions.  We can represent this by generalizing the basic model to
	\begin{equation}
	\begin{split}
	\dot{n}_A &= (1-s)(p+n_A) n_{AB} - n_A n_B, \\
	\dot{n}_B &= (1-s)n_B n_{AB} - (p+n_A) n_B, 
	\end{split}
	\label{eq2}
	\end{equation}
where the stubbornness parameter $s$ indicates how likely a moderate is to remain moderate after listening to an extremist. When $s$ is zero, we recover the basic model. 

Intuitively, one might expect that increasing $s$ should increase the size of the moderate subpopulation. Indeed, when $s$ is small enough, the moderate subpopulation does grow slightly with increasing $s$~[Figs.~\ref{GeneralizationsOnRealNetworks}(a) and~\ref{GeneralizationsOnRealNetworks}(d)].  But remarkably, if $s$ increases past a certain threshold, the moderates are driven to extinction; the size of their subpopulation drops to zero.  

We can examine this surprising behavior in another way by calculating how $s$ affects $p_c$ (the critical fraction of zealots needed for the revolution to succeed).  Intuition would suggest that $p_c$ should increase with $s$; the more stubborn the moderates are, the more zealots are needed to persuade them and everyone else.  In fact the opposite is true:  $p_c$ decreases with $s$, dropping monotonically from $1-\sqrt{3}/2$ at $s = 0$ to zero at $s = 1$~\cite{suppmat}.  Thus, increasing the stubbornness of the moderates makes the population \textit{more vulnerable} to takeover by the zealots. 

To make sense of why $p_c$ should decrease with $s$, it helps to realize that increasing $s$ not only reduces the flow of $AB$ individuals to opinion $A$ but also to opinion $B$, thereby depleting \textit{both} the uncommitted $A$ and $B$ subpopulations.  With competition from $B$ extremists over the $AB$ subpopulation weakened as a result, it takes fewer $A$ zealots (and hence a lower $p_c$) to convert the moderates to the $A$ camp.

This explanation suggests that evangelism is an important force in the dynamics.  So as a second strategy, we might try having the moderates actively promote moderation via the following generalization:
	\begin{equation}
	\begin{split}
	\dot{n}_A &= (p+n_A)n_{AB} - n_A n_B - r n_A n_{AB}, \\
	\dot{n}_B &= n_B n_{AB} - (p+n_A)n_B - r n_B n_{AB},
	\end{split}
	\label{eq4}
	\end{equation}
where the new parameter $r$ is a nonnegative real number that reflects the intensity of the moderates' evangelism.

Again it may seem intuitively clear that the size of the moderate fraction should increase if the moderates start actively deradicalizing the population.  For $r$ up to unity, however, the outcome is similar to that of making the moderates more stubborn. Figures~\ref{GeneralizationsOnRealNetworks}(b) and~\ref{GeneralizationsOnRealNetworks}(e) show that at a certain value of $r$, the size of the moderate subpopulation snaps discontinuously to zero.  If the moderates' campaign of persuasion is sufficiently successful from the start---that is, if $r$ starts and stays large enough---then the moderates do in fact maintain a large, robust equilibrium population.  However, if they fail to sustain this level of persuasiveness indefinitely, their evangelistic efforts can instigate their own extinction.

Finally, let us consider a third strategy:  suppose that the fanatics are deradicalized by a promoderation media campaign or other environmental stimulus rather than through social interaction with moderates.  We could then expect the dynamics to take the following form:
	\begin{equation}
	\begin{split}
	\dot{n}_A &= (p+n_A)n_{AB} - n_A n_B - u n_A, \\
	\dot{n}_B &= n_B n_{AB} - (p+n_A)n_B - u n_B,
	\end{split}
	\label{eq6}
	\end{equation}
where $u$ is a nonnegative parameter representing the rate at which the radicals abandon their radical position in response to the nonsocial stimulus.

In contrast to the first two strategies (as well as four others treated in the Supplemental Material~\cite{suppmat}), increasing the new parameter ($u$) in this system generally increases the equilibrium $n_{AB}$ toward a limit of $1-p$, with the one exception of a discontinuous drop partway through the ascent in Fig.~\ref{GeneralizationsOnRealNetworks}(c).  However, the drop is not to zero as it was for the other strategies, and it vanishes in the limit of small $p$.  Furthermore, following the drop, regrowth of $n_{AB}$ is rapid.  Hence, this mechanism of promoting moderation, which we might call \textit{nonsocial deradicalization}, provides the first acceptable means that we have found for expanding the moderate population in the midst of an ideological revolution.  This holds for the three strategies in this Letter, and also for the four others in the Supplemental Material~\cite{suppmat}.  

By itself, this final assessment should be regarded with caution.  We suggest that a greater emphasis be placed on our general approach as a framework for testing possible strategies as part of a continuing research program, which through further study might well uncover other means of fostering moderation more sophisticated than those considered here.

\newpage

\begin{acknowledgments}
H.~Hong acknowledges the hospitality of Cornell University during her visit for a sabbatical year.  This research was also supported in part by NSF Grants No.~CCF-0835706 and No.~CCF-0832782 (S.~H.~S.), and by the Michigan Society of Fellows (S.~A.~M.).
\end{acknowledgments}


\begin{thebibliography}{26}
\bibitem[{dif()}]{difference}
\bibinfo{note}{To our knowledge, the only other model in literature with a
  committed fraction and three opinion states is in Ref.~\cite{xie11}. This
  model is based on the Naming Game~\cite{castellano09}, and so has a
  substantially different list of interactions than our basic model. For
  example, it assumes that if both the speaker and listener are moderate, then
  they will spontaneously persuade each other to adopt one or the other extreme
  position. By contrast, our basic model regards this outcome as infrequent
  relative to radicalization when the speaker is extremist rather than
  moderate. For other models involving committed individuals (or similar
  entities), see Refs.~\cite{centola05, galam07}. Additional models with three
  opinion states are proposed in Refs.~\cite{vazquez03, vazquez04, gekle05,
  delalama06, castello06}.}
\end{thebibliography}
\end{document}


\title{Supplemental Material}

\maketitle

Here we derive the mean-field results described in the main text.  Our approach consists of analyzing seven different one-parameter generalizations of the basic model.  Our central conclusion is that only one of these, nonsocial deradicalization, offers a robust means of expanding the moderate subpopulation.  We finish with an example of a two-parameter generalization that, like nonsocial deradicalization, can deradicalize all uncommitted extremists without risking the extinction of the moderates.

In order to help interpret the results below, we recommend that the reader plot the nullclines of each system and study how they vary with the parameter values using the MATLAB program \texttt{pplane}.  This program is available online at \href{http://math.rice.edu/~dfield/dfpp.html}{http://math.rice.edu/$\sim$dfield/dfpp.html} and is free for educational use.

\section{The Basic Model}

The dynamics of the basic model are
	\begin{equation*}
	\begin{split}
	\dot{n}_A &= (p+n_A)n_{AB} - n_A n_B, \\
	\dot{n}_B &= n_B n_{AB} - (p+n_A)n_B,
	\end{split}
	\tag{1}
	\label{eq1}
	\end{equation*}
where the overdot represents differentiation by time and $n_{AB} = 1 - p - n_A - n_B$ (see the main text for definitions of the remaining terms).  The nullclines of (\ref{eq1}) are
	\begin{align*}
	0 &= (p+n_A)n_{AB} - n_A n_B, \tag{2a}\label{eq2a} \\
	0 &= n_B n_{AB} - (p+n_A)n_B. \tag{2b}\label{eq2b}
	\end{align*}
The first implies that
	\begin{equation*}
	n_B = \frac{(p+n_A)(1-p-n_A)}{p+2n_A},
	\tag{3}
	\label{eq3}
	\end{equation*}
and the second (by factoring) implies either $n_B = 0$ or
	\begin{equation*}
	n_B = 1 - 2p - 2n_A.
	\tag{4}
	\label{eq4}
	\end{equation*}

If we substitute $n_B = 0$ into (\ref{eq2a}), we obtain two solutions:  $n_A = -p$ and $n_A = 1-p$.  So (\ref{eq1}) has fixed points at $(n_A,n_B) = (-p,0)$ and $(n_A,n_B) = (1-p,0)$.

To obtain the remaining fixed points, we set (\ref{eq3}) and (\ref{eq4}) equal to each other and solve the resulting quadratic in $n_A$ to obtain
	\begin{equation*}
	n_A =  \tfrac{1}{6}\Bigl(1 - 4p \pm \sqrt{4p^2-8p+1}\Bigr).
	\tag{5}
	\label{eq5}
	\end{equation*}
Together with (\ref{eq4}), this gives a third and fourth fixed points so long as the discriminant of (\ref{eq5}) is nonnegative.  

Solving for the roots of the discriminant of (\ref{eq5}) gives $p_c = 1 \pm \sqrt{3}/2$.  We can disregard the value with the positive sign because $p$ represents a fraction of the total population and thus must be less than one.  The value with the negative sign represents the $p$ at which the third and fourth fixed points coalesce in a saddle-node bifurcation in $(n_A,n_B)$ phase space.

\section{Generalization I: \protect\\ Stubborn Moderates}

One way to encourage expansion of the moderate subpopulation would be to try to make existing moderates less likely to abandon their moderate views and convert to an extreme position.  This idea can be captured with the following generalization of (\ref{eq1}):
	\begin{equation*}
	\begin{split}
	\dot{n}_A &= (1-s)(p+n_A)n_{AB} - n_A n_B, \\
	\dot{n}_B &= (1-s)n_B n_{AB} - (p+n_A)n_B. 
	\end{split}
	\tag{6}
	\label{eq6}
	\end{equation*}
Here $s$ is a parameter on the unit interval.  Increases in $s$ correspond to the moderates becoming less prone to switch to a radical position.

Following the approach of the previous section, we find that this system has fixed points at $(n_A,n_B) = (-p,0)$ and $(n_A,n_B) = (1-p,0)$, as well as where
	\begin{equation*}
	n_A = \frac{(1-s)(1-3p)-p \pm \sqrt{\Delta(s,p)}}{6-4s}
	\tag{7}
	\label{eq7}
	\end{equation*}
and 
	\begin{equation*}
	n_B = \frac{(1-s)(1-p) - p - (2-s)n_A}{1-s},
	\tag{8}
	\label{eq8}
	\end{equation*}
with $\Delta(s,p) = (2-s)^2p^2-2(1-s)(4-3s)p+(1-s)^2$.  This second pair of fixed points exist everywhere that $\Delta(s,p) > 0$, and the curve $\Delta(s,p) = 0$ represents the parameter pairs $(s,p)$ at which the third and fourth fixed points coalesce in a saddle-node bifurcation.  

Solving $\Delta(s,p) = 0$ in terms of $p$ (i.e.~finding the roots of the quadratic polynomial in $p$) gives
	\begin{equation*}
	p_c = \frac{(1-s)(4-3s) \pm 2\sqrt{(1-s)^3(3-2s)}}{(2-s)^2}.
	\tag{9} 
	\label{eq9}
	\end{equation*}
(The $c$ subscript is added to indicate a connection to literature on critical phenomena and phase transitions.)  The root in (\ref{eq9}) with the plus sign represents a saddle-node bifurcation that occurs in the third quadrant of the $(n_A,n_B)$ plane where $n_A$ and $n_B$ are both negative, so the root with the minus sign is the only one of interest to us.  This curve descends monotonically from $p = 1-\sqrt{3}/2$ at $s = 0$ to $p = 0$ at $s = 1$.  For the $(s,p)$ pairs below it, $\Delta(s,p) > 0$, so the third and fourth fixed points exist for these $(s,p)$.  The third and fourth fixed points generally have small values for $n_A$ and large values for $n_B$, and the one having the $n_A$ with the negative sign in (\ref{eq7}) is stable.  (We can show this by an analysis of the Jacobian of (\ref{eq6}).)  By contrast for the $(s,p)$ pairs above the curve, $\Delta(s,p) < 0$, so the only fixed point of relevance in this region of parameter space is $(n_A,n_B) = (1-p,0)$.  Again by stability analysis on the Jacobian of (\ref{eq6}), this point is stable for all $(s,p)$ pairs on the unit square.

\section{Generalization II: \protect\\ Evangelical Moderates}

Another approach to increasing the moderate fraction would be to make the moderates evangelical for their moderate perspective.  To reflect this idea, we can generalize the basic model in the following way:
	\begin{equation*}
	\begin{split}
	\dot{n}_A &= (p+n_A)n_{AB} - n_A n_B - r n_A n_{AB}, \\
	\dot{n}_B &= n_B n_{AB} - (p+n_A)n_B - r n_B n_{AB},
	\end{split}
	\tag{10}
	\label{eq10}
	\end{equation*}
where $r$ is a nonnegative parameter representing the effectiveness of the moderates' evangelism.

Analyzing this model as we did the basic model, we again find that there are two fixed points on the $n_A$ axis:  $(n_A,n_B) = (1-p,0)$ and $(n_A,n_B) = (-p/(1-r),0)$, and two fixed points off the $n_A$ axis with locations given by
	\begin{equation*}
	n_A = \frac{(1-r)(1-3p)-(1+r)p \pm \! \sqrt{\Delta(r,p)}}{6-4r}
	\tag{11}
	\label{eq11}
	\end{equation*}
and 
	\begin{equation*}
	n_B = \frac{(1-r)(1-p) - p - (2-r)n_A}{1-r},
	\tag{12}
	\label{eq12}
	\end{equation*}
where $\Delta(r,p) = 4(1-r)^2p^2 - 4(1-r)(2-r)p + (1-r)^2$.  

The curve $\Delta(r,p) = 0$ gives the $(r,p)$ pairs at which a saddle-node bifurcation of the third and fourth fixed points occurs.  However since $p$ must be on the unit interval and $r$ is nonnegative, it can be easily shown that the only relevant segment of this curve is
	\begin{equation*}
	p_c = \frac{2-r - \sqrt{3-2r}}{2-2r},
	\tag{13}
	\label{eq13}
	\end{equation*}
where $r$ is restricted to the unit interval.  This segment starts at $p = 1-\sqrt{3}/2$ when $r = 0$ and declines monotonically to zero as $r$ reaches 1.  For allowed $(r,p)$ below the curve, $\Delta(r,p) > 0$.  So the two fixed points given by (\ref{eq11}) and (\ref{eq12}) exist in this region.  For allowed $(r,p)$ above the curve (\ref{eq13}), $\Delta(r,p) < 0$ and so (\ref{eq11}) and (\ref{eq12}) do not exist in this region and the only fixed points are $(n_A,n_B) = (1-p,0)$ and $(n_A,n_B) = (-p/(1-r),0)$.

There is also a second bifurcation, this time transcritical, that occurs when the fixed point on the $n_A$ axis located at $n_A = -p/(1-r)$ passes leftward through the fixed point at $n_A = 1-p$ as $r$ is increased through $r = 1/(1-p)$.  For comparison with (\ref{eq13}), we can write this bifurcation curve as a critical $p$:  $p_c = 1-1/r$.  This curve starts from $p = 0$ at $r = 1$ and asymptotically approaches $p = 1$ as $r$ is increased toward positive infinity.

From analysis of the Jacobian of (\ref{eq10}), we find that the fixed point with the $n_A$ in (\ref{eq11}) having the negative sign is stable, while the other is unstable.  We also find that the fixed point at $(n_A,n_B) = (1-p,0)$ is stable when $0 < r < 1/(1-p)$.  However as $r$ is increased through the transcritical bifurcation, the fixed point at $n_A = -p/(1-r)$ inherits the stability of the fixed point at $1-p$.  Thus for $r > 1/(1-p)$, the equilibrium value of $n_A$ decreases to zero like $1/r$ for large $r$.

\section{Generalization III: \protect\\ Nonsocial Deradicalization}

A third way to try to increase the size and stability of the moderate subpopulation would be to provoke radicals on both sides to become moderate by some \textit{nonsocial} means.  This strategy is captured by the following generalization of the basic model:
	\begin{equation*}
	\begin{split}
	\dot{n}_A &= (p+n_A)n_{AB} - n_A n_B - u n_A, \\
	\dot{n}_B &= n_B n_{AB} - (p+n_A)n_B - u n_B,
	\end{split}
	\tag{14}
	\label{eq14}
	\end{equation*}
where $u$ is a nonnegative parameter representing the rate of deradicalization.

We can again treat this system in the same way that we did the basic model.  The first pair of fixed points have $n_B = 0$ and
	\begin{equation*}
	n_A =  \tfrac{1}{2}\Bigl(1-2p-u \pm \sqrt{4up + (1-u)^2}\Bigr).
	\tag{15}
	\label{eq15}
	\end{equation*}
Note that the discriminant of (\ref{eq15}) is always positive, so these first two fixed points exist for all allowed $p$ and $u$.  For fixed $p$, the positive root of (\ref{eq15}) shrinks to zero like $1/u$ as $u$ is increased toward positive infinity.  This is easily seen from the asymptotic expansion of $n_A$ in powers of $1/u$, which begins
	\begin{equation*}
	n_A = p(1-p)\left(\frac{1}{u} + \frac{(1-2p)}{u^2} + \frac{(1-2p)^2}{u^3} + \cdots \right).
	\tag{16}
	\label{eq16}
	\end{equation*}
	
For the second pair of fixed points, the first coordinate is given by
	\begin{equation*}
	n_A =  \tfrac{1}{6}\Bigl(1-4p-u \pm \sqrt{\Delta(u,p)}\Bigr)
	\tag{17}
	\label{eq17}
	\end{equation*}
and the second coordinate by
	\begin{equation*}
	n_B = 1-2p-u - 2n_A,
	\tag{18}
	\label{eq18}
	\end{equation*}
where $\Delta(u,p) = 4p^2 - 4(2+u)p + (1-u)^2$.  

Solving $\Delta(u,p) = 0$ for $p$, we obtain the curve on which the saddle-node bifurcations occur:
	\begin{equation*}
	p_c = 1+\frac{u}{2} \pm \frac{\sqrt{3+6u}}{2}.
	\tag{19}
	\label{eq19}
	\end{equation*}
Only the root in (\ref{eq19}) with the minus sign is of interest to us, because only this root has values less than 1 for positive $u$.  Additionally, we are only interested in the segment of this curve with $u$ on the unit interval, since for larger $u$ the curve represents a saddle-node bifurcation in the third quadrant of the $(n_A,n_B)$ plane.

Below this curve segment, we find that $\Delta(u,p) > 0$, so the third and fourth fixed points exist there.  By analyzing the Jacobian of (\ref{eq14}), we can show that the fixed point having the $n_A$ in (\ref{eq14}) with the negative sign is stable and draws in trajectories to a small $n_A$ and large $n_B$.  However above the curve segment, we find that $\Delta(u,p) < 0$, so the only fixed points of (\ref{eq14}) are the two with $n_B = 0$ and $n_A$ values given by (\ref{eq15}).  The one of these with the positive $n_A$ value is stable and draws in all trajectories of interest for all allowed $(u,p)$ above the curve.

\section{Generalizations IV, V, and VI: 
\protect\\ Asymmetric Versions of Generalizations I, II, and III}

Generalizations I and II above lead to an annihilation of the moderate subpopulation for some values of $s$ and $r$.  This occurs because at these values of $s$ and $r$, much of the $B$ subpopulation is absorbed into the moderate subpopulation early in the time series, which allows the $A$ believers to subsequently convert the entire population to their belief $A$.  Motivated by this observation, we now consider three variations on the above generalizations that all attempt to shift the dynamics in favor of the $B$ believers.  We first show that each of these models has a fixed point that converges to $(n_A,n_B) = (0,1-2p)$ in a large-parameter limit.  We then discuss the implications of this for the fraction of moderates in each model.

To prepare the first variation (Generalization IV), we omit the factor of $(1-s)$ in the second equation of (\ref{eq6}) to obtain
	\begin{equation*}
	\begin{split}
	\dot{n}_A &= (1-s)(p+n_A)n_{AB} - n_A n_B, \\
	\dot{n}_B &= n_B n_{AB} - (p+n_A)n_B. 
	\end{split}
	\tag{20}
	\label{eq20}
	\end{equation*}
Like the original system, this new system has fixed points at $(n_A,n_B) = (-p,0)$ and $(n_A,n_B) = (1-p,0)$.  It also has fixed points where
	\begin{equation*}
	n_A = \frac{1-4p+2sp \pm \sqrt{4p^2-4(2-s)p+1}}{6-2s}
	\tag{21}
	\label{eq21}
	\end{equation*}
and $n_B = 1-2p-2n_A$.  In the limit as $s$ tends to 1, this second pair of points converges to $(n_A,n_B) = (1/2-p,0)$ and $(n_A,n_B) = (0,1-2p)$.

A second variation (Generalization V) is obtained by discarding the last term in the second equation of (\ref{eq10}).  This yields
	\begin{equation*}
	\begin{split}
	\dot{n}_A &= (p+n_A)n_{AB} - n_A n_B - r n_A n_{AB}, \\
	\dot{n}_B &= n_B n_{AB} - (p+n_A)n_B.
	\end{split}
	\tag{22}
	\label{eq22}
	\end{equation*}
The fixed points of this system are located at $(n_A,n_B) = (-p/(1-r),0)$ and $(n_A,n_B) = (1-p,0)$, and where
	\begin{equation*}
	n_A = \frac{1-4p+rp \pm \sqrt{(2-r)^2p^2-2(4-r)p+1}}{6-2s}
	\tag{23}
	\label{eq23}
	\end{equation*}
and $n_B = 1-2p-2n_A$.  In the large-$r$ limit, the fixed point having the $n_A$ with the minus sign in (\ref{eq23}) becomes $(n_A,n_B) = (0,1-2p)$.

Finally, the third variation (Generalization VI) involves omitting the final term in the second equation of (\ref{eq14}) to obtain
	\begin{equation*}
	\begin{split}
	\dot{n}_A &= (p+n_A)n_{AB} - n_A n_B - u n_A, \\
	\dot{n}_B &= n_B n_{AB} - (p+n_A)n_B.
	\end{split}
	\tag{24}
	\label{eq24}
	\end{equation*}
This system has two fixed points along the $n_A$ axis at
	\begin{equation*}
	n_A = \tfrac{1}{2}\Bigl(1-2p-u \pm \sqrt{4up+(1-u)^2}\Bigr).
	\tag{25}
	\label{eq25}
	\end{equation*}
It also has an additional two where
	\begin{equation*}
	n_A = \tfrac{1}{6}\Bigl(1-4p+u \pm \sqrt{4p^2-8(1+u)p+(1+u)^2}\Bigr),
	\tag{26}
	\label{eq26}
	\end{equation*}
and $n_B = 1-2p-2n_A$.  Analogous to the two cases above, the fixed point with the negative sign in (\ref{eq26}) converges to $(n_A,n_B) = (0,1-2p)$ in the large-$u$ limit.

The point $(n_A,n_B) = (0,1-2p)$ therefore appears as the limit of a fixed point in all three of these new models.  Furthermore at this fixed point, the moderate subpopulation is $n_{AB} = p$, so neither $p$ nor $n_{AB}$ can exceed 1/2 at this fixed point.  A more extensive analysis of the existence and stability of all fixed points shows that when $p < 1/2$ and each parameter ($s$, $r$, and $u$) is large enough, this fixed point exists in the first quadrant and is stable, receiving the trajectories of a large basin of attraction around it including the initial point $(n_A,n_B) = (0,1-p)$.

\section{Generalization VII: \protect\\ Opposing Zealots}

A seventh strategy for expanding the fraction of moderates in the population is to assemble a faction committed to $B$ that could counterbalance the existing faction committed to $A$.  If we represent the size of this second fraction by $q$, then the basic model becomes
	\begin{equation*}
	\begin{split}
	\dot{n}_A &= (p+n_A)n_{AB} - (q+n_B)n_A, \\
  \dot{n}_B &= (q+n_B)n_{AB} - (p+n_A)n_B,
	\end{split}
	\tag{27}
	\label{eq27}
	\end{equation*}
where now $n_{AB} = 1-p-q-n_A-n_B$.  

We can find exact expressions for the fixed points of this system, but they are more complicated than those of the first three generalizations above.  This is primarily due to the fact that we cannot factor out an $n_B$ from the left-hand side of the second rate equation in (\ref{eq27}), so we must use the quartic equation (rather than the quadratic equation) to obtain the fixed points.  Due to the size of the solutions, we don't include them here.  However they can be easily found with a computational platform for symbolic algebra such as Mathematica.

When we arbitrarily vary $p$ and $q$, the fixed points of (\ref{eq27}) generally undergo saddle-node bifurcations in the $(n_A,n_B)$ phase plane.  In the special case that we set $q = p$, varying the combined parameter (which we will just call $p$ from now on) produces a two-dimensional supercritical pitchfork bifurcation.  The handle and middle tine of this pitchfork lie along the line $n_B = n_A$, with the origin located along the handle.  At some critical $p$, the two outside tines of the pitchfork emerge and grow outward symmetrically about $n_B = n_A$.

To find the critical $p$ where this bifurcation occurs, let's consider the nullclines of (\ref{eq14}) under the substitution $q = p$:
	\begin{align*}
	0 &= (p+n_A)n_{AB} - (p+n_B)n_A, \tag{28a}\label{eq28a} \\
	0 &= (p+n_B)n_{AB} - (p+n_A)n_B. \tag{28b}\label{eq28b}
	\end{align*}
If we subtract each side of (\ref{eq28b}) from the corresponding side of (\ref{eq28a}), we obtain $0 = (n_A-n_B)(1-3p-n_A-n_B)$, so either $n_B = n_A$ or $n_A + n_B = 1-3p$.  The former, by substitution into (\ref{eq28a}), gives the fixed points on the line $n_B = n_A$:  $(n_A,n_B) = (-p,-p)$ and $(n_A,n_B) = ((1-2p)/3,(1-2p)/3)$.  The latter, also by substitution into (\ref{eq28a}), yields $0 = n_A^2 - (1-3p)n_A + p^2$, or
	\begin{equation*}
	n_A =  \tfrac{1}{2}\Bigl(1-3p \pm \sqrt{(1-3p)^2-4p^2}\Bigr).
	\tag{29}
	\label{eq29}
	\end{equation*}
Since both $p$ and $1-3p$ must be positive (recall that $1-3p$ is equal to $n_A + n_B$), the discriminant vanishes when $1-3p = 2p$, or when $p = 1/5$.  So this is the $p$ of the pitchfork bifurcation that we seek.

In fact we not only can find the $p$ of the pitchfork bifurcation but also the $(q,p)$ pairs at which the saddle-node bifurcations occur.  We can do so by identifying the parts of the fixed points of (\ref{eq27}) that are under even-powered roots.  In fact, there is only one such part (although it appears many times throughout the expressions for the fixed points) and setting it equal to zero gives an implicit function for the curve in parameter space along which the bifurcations occur.  With some simplification, this is
	\begin{equation*}
	\begin{split}
	32(p^3q&+pq^3) + 61p^2q^2 - 100(p^2q+pq^2) \\
	       &+ 82pq + 4(p^2+q^2) - 8(p+q) + 1 = 0.
	\end{split}
	\tag{30}
	\label{eq30}
	\end{equation*}
Since the highest power of $p$ and $q$ in (\ref{eq30}) is 3, this level set could instead be expressed as a set of two explicit functions, either $p$ in terms of $q$ or the other way around.  The two separate functions correspond to two different saddle-node bifurcations, and the locus where they meet marks the $p$ of the pitchfork bifurcation.

More broadly, (\ref{eq30}) represents a cusp catastrophe and can be shifted and rotated into a more conventional orientation (cusp positioned at the origin, pointing to the right) with the new parameters $a = p + q - 2/5$ and $b = p - q$.  The variable coordinate, orthogonal to the parameter plane, may be taken to be $n_A$, $n_B$, or some reasonable combination of these, for example $(n_A - n_B)/\sqrt{2}$ (the projection onto the line $0 = n_A + n_B$).

A variety of interesting results can be proven for the system (\ref{eq27}).  For example, one is that the $n_{AB}$ fraction at equilibrium cannot exceed $(3-\sqrt{3})/6$ ($< 0.212$).  Showing this consists of demonstrating that in either of the two regions of parameter space partitioned by the curve (\ref{eq30}), the fraction of $n_{AB}$ at equilibrium is greatest along this curve itself, with the maximum for the region with small $p$ and $q$ occurring at the intersection of (\ref{eq30}) with the $p$ axis, that is $(q,p) = (0,1-\sqrt{3}/2)$, and the maximum for the complementary region of parameter space occurring at the locus of the cusp bifurcation where $(q,p) = (1/5,1/5)$.  (Here we have assumed that the system is started from $(n_A,n_B) = (0,1-p)$, but the result holds for any initial condition.)  Evaluation of the function for $n_{AB}$ at these points demonstrates that the former has the larger value at $(3-\sqrt{3})/6$.

\newpage

\section{Generalization VIII: \protect\\ Stubborn Moderates \protect\\ and Opposing Zealots}

Finally, we analyze a model that combines Generalizations I and VII:
	\begin{equation*}
	\begin{split}
	\dot{n}_A &= (1-s)(p+n_A)n_{AB} - (q+n_B)n_A, \\
  \dot{n}_B &= (1-s)(q+n_B)n_{AB} - (p+n_A)n_B,
	\end{split}
	\tag{31}
	\label{eq31}
	\end{equation*}
where $n_{AB}$ is again $1-p-q-n_A-n_B$.  

Our aim in studying this model is to evaluate whether we can expand the moderate subpopulation with the following two-part strategy:  (i) assemble a faction committed to $B$ of the same size as that committed to $A$, and (ii) increase $s$ (the stubbornness of the moderates to remain moderate).  For simplicity, we suppose that we have already rallied a faction committed to $B$ equal to the size of that committed to $A$, or, in the parlance of the model, $q = p$.

This leaves us with the following nullclines:
	\begin{align*}
	0 &= (1-s)(p+n_A)n_{AB} - (p+n_B)n_A, \tag{32a}\label{eq32a} \\
	0 &= (1-s)(p+n_B)n_{AB} - (p+n_A)n_B. \tag{32b}\label{eq32b}
	\end{align*}
We expect a pitchfork bifurcation to occur at some value $p$ as a function of $s$.  For $s = 0$, this $p$ is just 1/5, as we found in our analysis of Generalization VII.

If we subtract each side of (\ref{eq32b}) from the corresponding side of (\ref{eq32a}), we obtain $0 = (n_A-n_B)[(1-s)(1-2p-n_A-n_B)-p]$, so either $n_B = n_A$ or $n_A + n_B = 1-2p-p/(1-s)$.  The former combined with (\ref{eq32a}) gives the fixed points $n_A = n_B = -p$ and $n_A = n_B = (1-s)(1-2p)/(3-2s)$, while the latter combined with same equation yields $0 = n_A^2 - \lambda n_A + p^2$, where we have defined $\lambda = 1-2p-p/(1-s)$.  Thus we have
	\begin{equation*}
	n_A =  \tfrac{1}{2}\Bigl(\lambda \pm \sqrt{\lambda^2-4p^2}\Bigr).
	\tag{33}
	\label{eq33}
	\end{equation*}
Now $p$ must be positive, and since $\lambda$ is the sum $n_A+n_B$, it also must be positive.  So the discriminant vanishes when $\lambda = 2p$, or at
	\begin{equation*}
	p_c = \frac{1-s}{5-4s},
	\tag{34}
	\label{eq34}
	\end{equation*}
which is the $p$ at which the pitchfork bifurcation occurs.  Note also that at this $p_c$ we have the beautiful result that $n_A = n_B = p$.

Since (\ref{eq34}) starts at $p = 1/5$ when $s = 0$ and descends monotonically to $p = 0$ at $s = 1$, we can see immediately that no matter how small $p$ is, if we increase $s$ enough we will cross through the $p_c$ given by (\ref{eq34}) and obtain an $n_{AB}$ of $(1-2p)/(3-2s)$.  Moreover, before crossing this $p_c$, $n_{AB}$ is of size $p/(1-s)$.  With additional work, we can show that this implies the rise in $n_{AB}$ is continuous and therefore robust to minor variations in $s$ and $p$.

\begin{figure*}[h!]
\includegraphics[scale = 1]{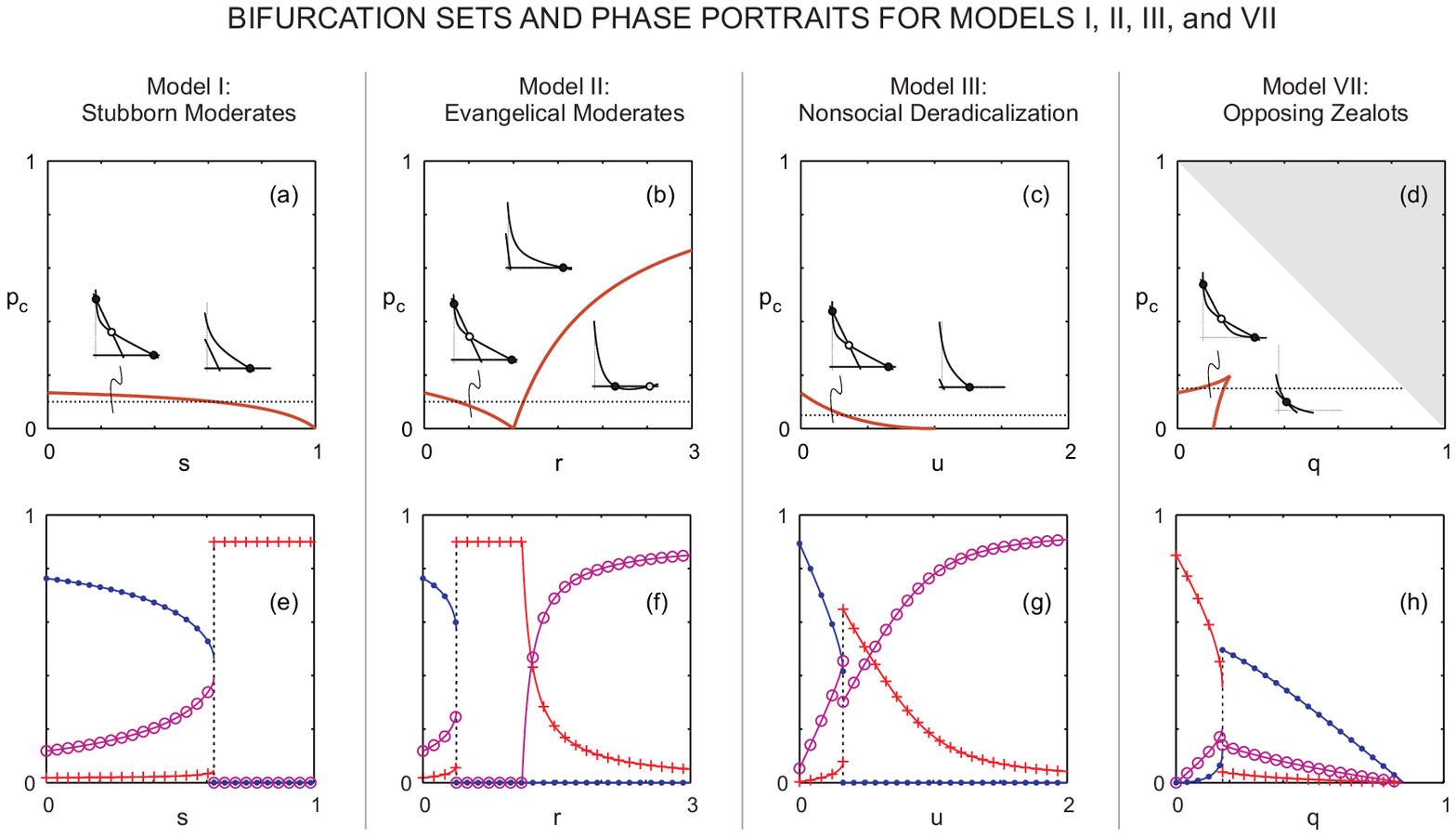}
\caption{(color online) (a)-(d):~Bifurcation sets for the models I, II, III, and VII with representative phase portraits shown in miniature.  The solid brown curves indicate boundaries along which bifurcations occur.  Specifically, the curves in (a) and (c), each branch of the curve in (d), and the left curve in (b) mark parameter pairs at which saddle-node bifurcations occur.  The right curve in (b) signifies a transcritical bifurcation, and the cusp of the curve in (d) coincides with a supercritical pitchfork bifurcation.  More broadly, the curve in (d) reflects a cusp catastrophe.  The qualitative structure of the phase plane changes across these bifurcation curves.  However it remains the same within each connected region and is illustrated by the inset phase portraits.  Each phase portrait has horizontal and vertical axes of $n_A$ and $n_B$ respectively and is restricted to the unit square.  The solid lines are nullclines and the closed and open dots represent, respectively, stable and unstable fixed points.  The horizontal dotted lines show the trajectories followed by corresponding plots (e)-(h).  From left to right, the $p$ values of these dotted lines are 0.1, 0.1, 0.05, and 0.15.
(e)-(h):~The final equilibrium values of $n_A$ (red plus signs), $n_B$ (blue dots) and $n_{AB}$ (magenta open circles) for the initial condition $(n_A,n_B) = (0,1-p)$ as a function of the new parameters ($s$, $r$, $u$, and $q$).  Note that only nonsocial deradicalization allows for arbitrary growth of the moderate population without risking its extinction.  All curves in panels (a)-(h) are computed exactly in the above sections.
\label{FourParameterPlanes}}
\end{figure*}